\documentclass[aps,prl,preprint,superscriptaddress]{revtex4}
\usepackage{graphicx}% Include figure files
\usepackage{pslatex}

\begin{document}

%\preprint{}

\title{Unit cell of graphene on Ru(0001): a 25 x 25
  supercell with 1250 carbon atoms}


\author{D.~Martoccia}

\affiliation{Swiss Light Source, Paul Scherrer Institut, CH-5232 Villigen, Switzerland.}

\author{P.R.~Willmott}
\email[]{philip.willmott@psi.ch}

\affiliation{Swiss Light Source, Paul Scherrer Institut, CH-5232 Villigen, Switzerland.}

\author{T. Brugger} 

\affiliation{Institute of Physics, University of Z\"urich,
  Winterthurerstrasse 190, CH-8057 Z\"urich, Switzerland.} 

\author{M. Bj\"orck}

\affiliation{Swiss Light Source, Paul Scherrer Institut, CH-5232 Villigen, Switzerland.}

\author{S. G\"unther}

\affiliation{Department of Chemistry, Universit\"at M\"unchen, D-81377
  M\"unchen, Germany.}

\author{C.M.~Schlep\"utz}

\affiliation{Swiss Light Source, Paul Scherrer Institut, CH-5232 Villigen, Switzerland.}

\author{A. Cervellino} 

\affiliation{Swiss Light Source, Paul Scherrer Institut, CH-5232 Villigen, Switzerland.}

\author{S.A. Pauli} 

\affiliation{Swiss Light Source, Paul Scherrer Institut, CH-5232 Villigen, Switzerland.}

\author{B.D.~Patterson}

\affiliation{Swiss Light Source, Paul Scherrer Institut, CH-5232 Villigen, Switzerland.}

\author{S. Marchini}

\affiliation{Department of Chemistry, Universit\"at M\"unchen, D-81377
  M\"unchen, Germany.}

\author{J. Wintterlin}

\affiliation{Department of Chemistry, Universit\"at M\"unchen, D-81377
  M\"unchen, Germany.}

\author{W. Moritz} 

\affiliation{Department of Crystallography, Universit\"at M\"unchen, D-81377
  M\"unchen, Germany.} 

\author{T. Greber}

\affiliation{Institute of Physics, University of Z\"urich,
  Winterthurerstrasse 190, CH-8057 Z\"urich, Switzerland.} 


\date{\today}

\begin{abstract} 
The structure of a single layer of graphene on Ru(0001) has been
studied using surface x-ray diffraction. A surprising superstructure
has been determined, whereby $25 \times 25$ graphene unit cells lie on
$23 \times 23$ unit cells of Ru. Each supercell contains $2
\times 2$ crystallographically inequivalent subcells caused by
corrugation. Strong intensity oscillations in the superstructure rods
demonstrate that the Ru substrate is also significantly corrugated
down to several monolayers, and that the bonding between 
graphene and Ru is strong and cannot be caused by
van der Waals bonds. Charge transfer from the Ru substrate to
the graphene expands and weakens the C--C bonds, which helps
accommodate the in-plane tensile stress. The elucidation of
this superstructure provides important information in the potential
application of graphene as a template for nanocluster arrays.
\end{abstract}


\newcommand{\etal}{{\em et al.}}
\newcommand{\degr}{$^{\rm o}$}
\newcommand{\degc}{$^{\rm o}$C}

\maketitle

%\section{Introduction}
%\label{sec:intro} 
% If in two-column mode, this environment will change to single-column
% format so that long equations can be displayed. Use
%\begin{widetext}
%\end{widetext}

The detailed structure determination of single-layer graphene on
well-defined surfaces is a significant goal in materials science and
solid-state physics -- it is most probable that future electronic
devices based on graphene layers will be fabricated on crystalline
substrates \cite{Keim07nmat}, hence a knowledge of how 
substrates affect graphene is of paramount importance
if the latter's structural and electronic properties are to be tailored 
\cite{Novoselov07}. In addition, it has been recently discovered
\cite{Marchini07,Coraux08,Pan07,Parga08,Sutter08} that, when grown on
crystalline transition metal surfaces, graphene can form
superstructures resulting from moir\'e superpositions of $(m~\times~m)$ 
carbon hexagons on $(n~\times~n)$ metal surface cells. 
It is still disputed as to whether observed features within these
supercells are caused by electron density fluctuations over a
basically flat structure \cite{Parga08}, or whether there is an 
actual buckling of the graphene sheet \cite{Marchini07,Wang08}. A structural
clarification would identify the potential of graphene in applications
such as molecular recognition, single-molecule
sensing \cite{Dil08}, and nanocluster array templates for biological
or catalytic
applications \cite{Min08nmat,Diaye06,Boyen02}. Here we show, 
using surface x-ray
diffraction (SXRD), that graphene forms a surprising superstructure when
grown on Ru(0001), whereby $25~\times~25$ graphene unit cells lie 
commensurately on $23~\times~23$ unit cells of Ru. Characteristic 
intensity oscillations in
the SXRD data prove that not only the graphene but also the 
Ru down to several atomic layers 
are significantly corrugated, indicating that the bonding between 
the single graphene layer and Ru is unusually strong.

The structure of graphene on Ru(0001) has already 
been investigated using scanning tunneling microscopy (STM),
conventional electron microscopy, x-ray photoelectron spectroscopy, 
Raman spectroscopy, and low-energy electron
diffraction (LEED) and microscopy
\cite{Wu94,Marchini07,Pan07,Parga08,Sutter08}, as well as density
functional theory (DFT) \cite{Wang08}. 
Until now it has remained a contentious issue as to what registry
exists between graphene and the underlying Ru substrate. It has been
suggested that $(12 \times 12)$ graphene hexagons sit on 
$(11 \times 11)$ Ru surface nets (described henceforth as 
$12$-on-$11$) \cite{Wu94,Marchini07,Pan07}, 
while an $11$-on-$10$ structure has also been proposed \cite{Parga08}. 
If one assumes an in-plane lattice constant for graphene equal 
to that for graphite ($2.4612$~\AA), the former structure would have 
an in-plane tensile strain on Ru ($a = 2.706$~\AA) 
of $0.78$~\%, while the latter would be 
compressively strained by only $0.05$~\%. 

The positions of the first order diffraction signals associated with
these two superstructures are $1.100$ and $1.0909$ 
in-plane reciprocal lattice units (r.l.u.) of the underlying Ru 
lattice, i.e., they lie within less than $0.01$~r.l.u. of one
another. With a conventional LEED system, one can only achieve 
accuracies of about one to two percent, excluding an unambiguous 
identification of the structure
using this method. Only surface x-ray diffraction is capable of 
achieving this goal \cite{Goriachko07,Feidenhansl89,Bunk07}. 

\begin{figure}
\includegraphics[scale=0.35]{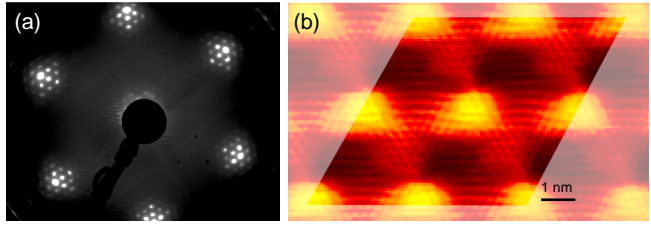}
\caption{\label{fig:leedstm} (color online) (a) A LEED image of the 
  graphene/Ru(0001) surface
  taken at an electron energy of $74$~eV. (b) An STM
  image of graphene on Ru(0001), highlighting the supercell containing four
  subcells.}
\end{figure}

% Experimental 
Two samples of 
graphene were prepared on separate occasions on the
same sputtered and annealed Ru(0001) 
single crystal. In both cases, the crystal was heated to 
$1115$~K in ultra-high vacuum
(UHV) and a single layer of graphene was deposited 
by dosing ethene at a pressure of $2 \times 10^{-5}$~Pa for
three minutes \cite{Oshima97}. The temperature was then held at 
$1115$~K for a further
$60$~seconds. For the first sample, the crystal was then cooled at a rate of
$0.4$~K\,s$^{-1}$ down to $915$~K, and from there at a rate of
$0.8$~K\,s$^{-1}$ down to $610$~K, after which the heating was turned
off. In the case of the second sample, cooling was approximately four
times quicker. It is noted here that these different cooling rates
were chosen to establish whether they affected the final form of the
absorbate structure. No significant difference between the two samples
could be detected, however. Typical LEED and STM images after cooling
to room temperature are shown in Fig.~\ref{fig:leedstm}. 

The samples were transferred in UHV to a minichamber ($10^{-7}$~Pa) 
equipped with a hemispherical Be-dome for studies using SXRD
\cite{ZegenhagenThanks}. The
chamber was mounted on the surface diffractometer of the Materials
Science beamline, Swiss Light Source \cite{Willmott05ass}. Structure
factors were recorded using the Pilatus 100k pixel detector
\cite{Schlepuetz05}. The photon energy was $12.4$~keV and the
transverse and longitudinal coherence lengths were both $1$~$\mu$m. 

%Results
\begin{figure*}
\includegraphics[scale=0.25]{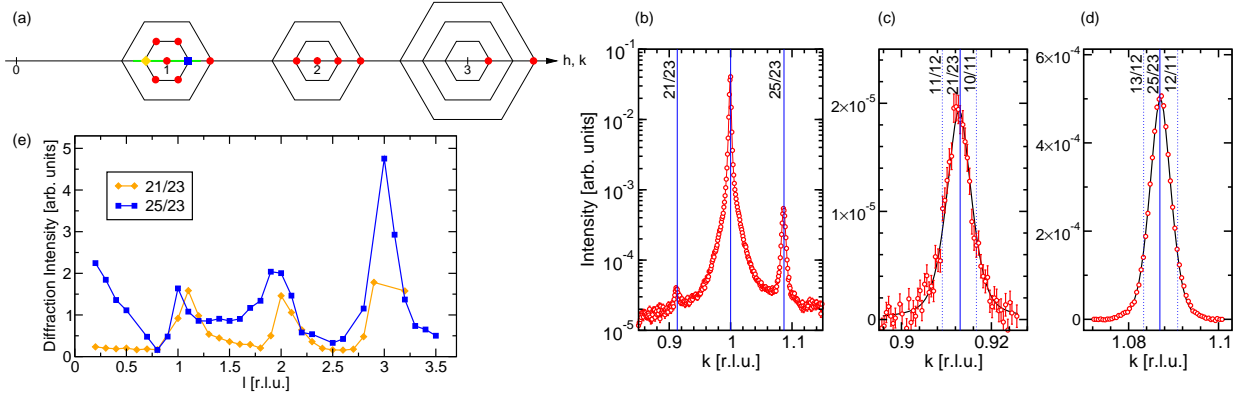}
\caption{\label{fig:diffrSummary} (color online) Summary of the
  diffraction data for the graphene/Ru(0001) system. (a)
  Schematic reciprocal space map showing where data was recorded. The red
  circular dots indicate points recorded in plane at $l = 0.4$~r.l.u. The
  in-plane scan along the
  $k$-direction in the neighborhood of the (01)~CTR of Ru at $l =
  0.4$~r.l.u. shown in (b) is indicated by the green (gray) line. The
  positions of superstructure rods shown in (e) are indicated by the 
  orange diamond and blue square. (c) and (d) High-resolution scans
  across the superstructure signal of graphene on Ru, detailing
  the $25$-on-$23$ reconstruction. The data were fit to a pseudo-Voigt
  profile (solid black curves), while the positions where 
  $13$-on-$12$ and $12$-on-$11$ reconstruction diffraction signals 
  would lie (dotted lines) are also shown.}
\end{figure*}

A calibration of reciprocal space was achieved to two parts
in $10\,000$ by using bulk Ru Bragg reflections as reference points. A
typical set of in-plane scans is shown in Fig.~\ref{fig:diffrSummary}. In
Fig.~\ref{fig:diffrSummary}(b), satellites on either side of the Ru~$(01)$
crystal truncation rod (CTR) can be seen. The primary graphene~$(01)$
signal [see Fig.~\ref{fig:diffrSummary}(d)] at $k = 1.087$~r.l.u. 
is, in itself, 
no proof of a commensurate reconstruction, as the graphene 
could in principle lie incommensurately above the Ru substrate. However, the
presence also of a signal an equal distance on the other side of the Ru~CTR
indicates that a reconstruction must exist
[Fig.~\ref{fig:diffrSummary}(c)]. We found several other signals proving a 
true reconstruction elsewhere in reciprocal space 
[see Fig.~\ref{fig:diffrSummary}(a)] Note that the (0001) surface of 
hexagonal close-packed systems  commonly display sixfold symmetry,
although the symmetry of a perfect hcp(0001) surface is
threefold. This apparent increase in symmetry is caused by surfaces
containing regions separated by atomic steps of half a unit cell
height, resulting in a $180^{\circ}$ rotation of adjacent
terraces. This is why only one in-plane axis is shown in
Fig.~\ref{fig:diffrSummary}(a). 

The positions of the two reconstruction signals shown in
Fig.~\ref{fig:diffrSummary}(c) and (d) and those of all the 
other superstructure signals investigated,
indicate however, that the superstructure complies with neither of
those proposed so far. In fact, the signals sit exactly at 
$21/23$ and $25/23$~r.l.u.,
to within $0.0002$~r.l.u. from which it is unambiguously clear from our SXRD
data that the reconstruction is in fact $25$-on-$23$, that is $25
\times 25$ graphene honeycombs sitting commensurately on $23 \times 23$ Ru unit
cells. This signal cannot be explained as 
originating from the incoherent addition of diffraction signals 
from large domains of $13$-on-$12$ and $12$-on-$11$ supercells, 
as the linewidth of the superstructure signal is, at $(5.4 \pm 0.1) \times
10^{-3}$~r.l.u., significantly narrower than the separation of any
independent $13$-on-$12$ and $12$-on-$11$ signals of $7.6 \times
10^{-3}$~r.l.u. [see Fig.~\ref{fig:diffrSummary}(c) and (d)]. Also, a
model consisting of a random distribution of $13$-on-$12$ and
$12$-on-$11$ supercells differs from that of the $25$-on-$23$
structure by maximum in-plane displacements of the carbon atoms of 
less than $0.1$~\AA, which are significantly smaller than
typical vibrational amplitudes at room temperature, hence can be
ignored. 

The large extent of this supercell, covering over $33$~nm$^2$ and
containing $1250$ carbon atoms is surprising, since the
structure forms in a process involving temperatures above $1000$~K, 
where more than $300$~eV thermal vibrational energy is stored in the
supercell. 
In the scanning tunneling microscope image of graphene on Ru(0001)
shown in Fig.~\ref{fig:leedstm}(b), a supercell is highlighted. This 
contains not one, but four parallelogram 
structures. It is still disputed whether the 
hill-like features are formed by a physical corrugation
of the graphene sheet, or is caused by electron density waves in an
essentially flat graphene layer \cite{Parga08}. Although this 
cannot be decided from the STM data
alone, a recent combined DFT/STM study \cite{Wang08} revealed that the graphene
is indeed significantly corrugated when deposited on Ru(0001). 
It is these features that make this
system so interesting as a potential nanotemplate. 

The four ``subcells'' within the supercell cannot map
translationally onto one another, as from the SXRD data it is clear
that the number of unit cells of Ru ($23$) as
well of graphene ($25$) along the edges of the 
supercell are {\em odd}, and one is therefore forced to conclude 
that the graphene supercell must 
consist of four translationally inequivalent subcells.  
It is noted that, because of the presence of $2 \times 2$ corrugation
periods within each supercell, all superstructure 
peaks with an in-plane distance from the Ru signals of $p/23$~r.l.u., 
where $p$ is an odd integer, are systematically absent. 

The $21/23$ and $25/23$ superstructure rods (SSRs) are shown in
Fig~\ref{fig:diffrSummary}(e). Because they only provide information
on surface reconstructions of the graphene and uppermost Ru-layers, 
and not on bulk properties, 
we are able to infer important properties of the surface region 
immediately from their qualitative features. First, the signal
intensity is strongly modulated, with a periodicity of approximately 
$1.0$~r.l.u. (with respect to Ru) in the
out-of-plane direction. This modulation can only occur if the
ruthenium is physically corrugated, that is, if the graphene imposes
vertical strain \cite{Wang08}. Importantly, all the maxima have widths 
of approximately $0.25$~r.l.u., which means that the Ru
substrate must also be significantly corrugated down
to about $4$ unit cells, or over $1.5$~nm. 

Unsurprisingly, the $25/23$ rod has significant intensity at low
$l$-values, as at $l=0$, this corresponds to the $(010)$ in-plane 
graphene peak, which is known to have non-zero intensity. However, 
extrapolation of the $21/23$-rod to $l=0$ strongly indicates that it 
also has non-zero intensity here, which can only occur if there are 
in-plane movements of the atoms within the supercell. 

\begin{figure}
\includegraphics[scale=0.335]{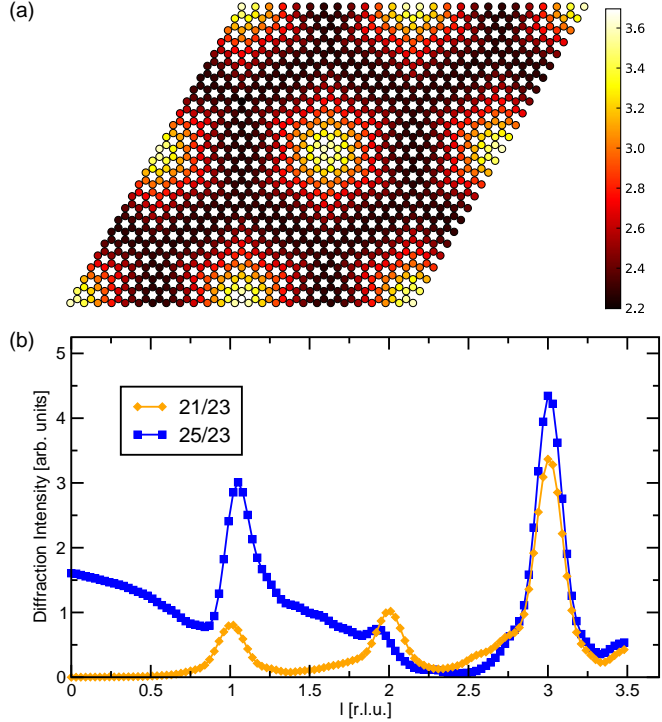}
\caption{\label{fig:simul} (color online) Simple parametric model of the
  graphene-Ru(0001) supercell. (a) The vertical displacement field of
  the graphene corrugation. The scale is given in \AA\
  above the Ru substrate. (b) Simulated $21/23$ and 
  $25/23$ superstructure rods of corrugated graphene on Ru(0001), 
  incorporating the qualitative
  features extracted from the experimental data. The graphene has a 
  peak-to-peak corrugation amplitude of $1.5$~\AA, 
  while that of the uppermost Ru
  atomic layer is $0.20$~\AA. The corrugation amplitude of the Ru
  drops exponentially by a factor of $0.75$ for each successive atomic
  layer ($0.56$ per unit cell depth). The minimum graphene--Ru
  distance is $2.2$~\AA, and the mean vertical positions between all
  the Ru atomic planes assume bulk values.}
\end{figure}

In order to estimate the corrugation amplitudes and depths of the 
ruthenium, we used a
simple model to simulate the two superstructure rods of
Fig~\ref{fig:diffrSummary}(e). The vertical displacement field of
the graphene corrugation [Fig.~\ref{fig:simul}(a)] is generated by
first calculating a subfield for each of the two inequivalent carbon
atoms in the conventional graphite unit cell, whereby the vertical
distance to the Ru-substrate is proportional to the in-plane
separation from the nearest Ru-atom. The component fields are
interpolated and then added to produce the final displacement map. The
model incorporated only four parameters, namely the graphene
corrugation amplitude and the minimum graphene-Ru distance, for which we
used fixed values determined by DFT calculations \cite{Wang08}, and 
the corrugation of the uppermost Ru atomic
layer and the exponential decay depth of the Ru-corrugation, which
were varied to best match the experimental linewidths and intensities. 
Any in-plane movements of either the graphene or
the Ru were ignored, as this would add significant complexity to the 
model, and our
a priori knowledge of such movements is very limited. The simulation
is shown in Fig.~\ref{fig:simul}. Despite the simplicity of this model, 
the qualitative agreement is impressive. The peak-to-peak 
corrugation amplitude 
of the uppermost ruthenium atomic layer is $0.20$~\AA, which decays 
exponentially with depth, with a characteristic length of $1.7$~unit 
cells ($3.4$~atomic layers). Note that 
the graphene and ruthenium corrugations were chosen to be in phase
(i.e., peak above peak and valley above valley). If the
corrugations are made to be out of phase (peak above valley, valley
above peak), the agreement with the
experimental data is poorer. 

The effect on the shape of
the SSRs of the graphene corrugation amplitude is fairly insensitive,
due to the low x-ray scattering amplitude of carbon. 
I(V) curves from low-energy
electron diffraction may provide important further quantitative
information regarding the detailed graphene structure, due to its 
higher surface sensitivity. 
A rigorous fit would also include in-plane
movements of both the carbon and Ru atoms and the possibility that 
the average height of each Ru atomic layer can vary. Allowing in-plane 
movements could significantly change the magnitudes of the 
calculated corrugations, hence their 
values given here are still tentative. 

The large depth to which the ruthenium substrate is perturbed is,
however, a robust parameter, and is 
indicative of an attendant strong chemical bonding of graphene to
the metallic substrate via the carbon $p_z$~orbitals, which cannot
arise through van der Waals bonds. This has recently been predicted by 
DFT calculations, that find strong charge redistributions and a minimum 
graphene-Ru distance of only $2.2$~\AA, incompatible with
van-der-Waals interactions \cite{Wang08}. Strong interactions have 
already been seen for graphene on Ni(111) \cite{Nagashima94,Oshima97}. This 
increased bonding to the substrate is connected with expanded and 
weakened C--C bonds, as indicated by the softened phonons of the 
graphene layer \cite{Wu94}. 

This in-plane expansion of the graphene layer might therefore
accommodate the apparent tensile strain of the $25$-on-$23$ supercell
when assuming a bulk-like graphite in-plane lattice
constant for graphene. Indeed, allowing for an expansion of the C--C
bond of approximately $1$~\% due to electron transfer would result in 
nominally zero
heteroepitaxial strain for this $25$-on-$23$ structure, although, of
course, because we observe elastic deformation of the Ru substrate,
there must be stress in the graphene layer that imposes strain both in
itself and in the substrate. In other
words, the Ru--C bonding causes the graphene to
dilate in plane and the $25$-on-$23$ structure to have the
lowest surface energy. 

In conclusion, the structure of the supercell of graphene on Ru(0001)
has been elucidated and been shown to consist of $25 \times 25$ unit
cells of graphene on $23 \times 23$ unit cells of Ru. On the one hand,
this large in-plane extent of over $6$~nm, and on the other, the large
corrugation amplitudes of the Ru-substrate, which indicate a strong bond
between the graphene and ruthenium, suggests that this system 
may be ideally suited as a robust template for arrays of nanoclusters or 
macromolecules. 

\begin{acknowledgments}
 The authors thank Dr. Marie-Laure Bocquet and Mr. Bin Wang 
 for fruitful discussions. Support of this work by the 
 Schweizerischer Nationalfonds zur
 F\"orderung der wissenschaftlichen Forschung and the staff of the
 Swiss Light Source is gratefully acknowledged. This work was partly
 performed at the Swiss Light Source, Paul Scherrer Institut. 
\end{acknowledgments}

\bibliography{graphene}

\begin{thebibliography}{21}
\expandafter\ifx\csname natexlab\endcsname\relax\def\natexlab#1{#1}\fi
\expandafter\ifx\csname bibnamefont\endcsname\relax
  \def\bibnamefont#1{#1}\fi
\expandafter\ifx\csname bibfnamefont\endcsname\relax
  \def\bibfnamefont#1{#1}\fi
\expandafter\ifx\csname citenamefont\endcsname\relax
  \def\citenamefont#1{#1}\fi
\expandafter\ifx\csname url\endcsname\relax
  \def\url#1{\texttt{#1}}\fi
\expandafter\ifx\csname urlprefix\endcsname\relax\def\urlprefix{URL }\fi
\providecommand{\bibinfo}[2]{#2}
\providecommand{\eprint}[2][]{\url{#2}}

\bibitem[{\citenamefont{Geim and Novoselov}(2007)}]{Keim07nmat}
\bibinfo{author}{\bibfnamefont{A.~K.} \bibnamefont{Geim}} \bibnamefont{and}
  \bibinfo{author}{\bibfnamefont{K.~S.} \bibnamefont{Novoselov}},
  \bibinfo{journal}{Nature Mater.} \textbf{\bibinfo{volume}{6}},
  \bibinfo{pages}{183} (\bibinfo{year}{2007}).

\bibitem[{\citenamefont{Novoselov}(2007)}]{Novoselov07}
\bibinfo{author}{\bibfnamefont{K.~S.} \bibnamefont{Novoselov}},
  \bibinfo{journal}{Nature Mater.} \textbf{\bibinfo{volume}{6}},
  \bibinfo{pages}{720} (\bibinfo{year}{2007}).

\bibitem[{\citenamefont{Marchini et~al.}(2007)\citenamefont{Marchini,
  G{\"{u}}nther, and Wintterlin}}]{Marchini07}
\bibinfo{author}{\bibfnamefont{S.}~\bibnamefont{Marchini}},
  \bibinfo{author}{\bibfnamefont{S.}~\bibnamefont{G{\"{u}}nther}},
  \bibnamefont{and}
  \bibinfo{author}{\bibfnamefont{J.}~\bibnamefont{Wintterlin}},
  \bibinfo{journal}{Phys. Rev. B} \textbf{\bibinfo{volume}{76}},
  \bibinfo{pages}{075429} (\bibinfo{year}{2007}).

\bibitem[{\citenamefont{Coraux et~al.}(2008)\citenamefont{Coraux, N'Diaye,
  Busse, and Michely}}]{Coraux08}
\bibinfo{author}{\bibfnamefont{J.}~\bibnamefont{Coraux}},
  \bibinfo{author}{\bibfnamefont{A.~T.} \bibnamefont{N'Diaye}},
  \bibinfo{author}{\bibfnamefont{C.}~\bibnamefont{Busse}}, \bibnamefont{and}
  \bibinfo{author}{\bibfnamefont{T.}~\bibnamefont{Michely}},
  \bibinfo{journal}{Nanolett.} \textbf{\bibinfo{volume}{8}},
  \bibinfo{pages}{565} (\bibinfo{year}{2008}).

\bibitem[{\citenamefont{Yi et~al.}(2007)\citenamefont{Yi, Dong-Xia, and
  Hong-Jun}}]{Pan07}
\bibinfo{author}{\bibfnamefont{P.}~\bibnamefont{Yi}},
  \bibinfo{author}{\bibfnamefont{S.}~\bibnamefont{Dong-Xia}}, \bibnamefont{and}
  \bibinfo{author}{\bibfnamefont{G.}~\bibnamefont{Hong-Jun}},
  \bibinfo{journal}{Chin. Phys.} \textbf{\bibinfo{volume}{16}},
  \bibinfo{pages}{3151} (\bibinfo{year}{2007}).

\bibitem[{\citenamefont{de~Parga et~al.}(2008)\citenamefont{de~Parga, Calleja,
  Borca, Passeggi, Hinarejos, Guinea, and Miranda}}]{Parga08}
\bibinfo{author}{\bibfnamefont{A.~L.~V.} \bibnamefont{de~Parga}},
  \bibinfo{author}{\bibfnamefont{F.}~\bibnamefont{Calleja}},
  \bibinfo{author}{\bibfnamefont{B.}~\bibnamefont{Borca}},
  \bibinfo{author}{\bibfnamefont{M.~C.~G.} \bibnamefont{Passeggi},
  \bibfnamefont{Jr.}}, \bibinfo{author}{\bibfnamefont{J.~J.}
  \bibnamefont{Hinarejos}},
  \bibinfo{author}{\bibfnamefont{F.}~\bibnamefont{Guinea}}, \bibnamefont{and}
  \bibinfo{author}{\bibfnamefont{R.}~\bibnamefont{Miranda}},
  \bibinfo{journal}{Phys. Rev. Lett.} \textbf{\bibinfo{volume}{100}},
  \bibinfo{pages}{056807} (\bibinfo{year}{2008}).

\bibitem[{\citenamefont{Sutter et~al.}(2008)\citenamefont{Sutter, Flege, and
  Sutter}}]{Sutter08}
\bibinfo{author}{\bibfnamefont{P.~W.} \bibnamefont{Sutter}},
  \bibinfo{author}{\bibfnamefont{J.-I.} \bibnamefont{Flege}}, \bibnamefont{and}
  \bibinfo{author}{\bibfnamefont{E.~A.} \bibnamefont{Sutter}},
  \bibinfo{journal}{Nature Mater.} \textbf{\bibinfo{volume}{7}},
  \bibinfo{pages}{406} (\bibinfo{year}{2008}).

\bibitem[{\citenamefont{Wang et~al.}(2008)\citenamefont{Wang, Bocquet,
  Marchini, G{\"{u}}nther, and Wintterlin}}]{Wang08}
\bibinfo{author}{\bibfnamefont{B.}~\bibnamefont{Wang}},
  \bibinfo{author}{\bibfnamefont{M.-L.} \bibnamefont{Bocquet}},
  \bibinfo{author}{\bibfnamefont{S.}~\bibnamefont{Marchini}},
  \bibinfo{author}{\bibfnamefont{S.}~\bibnamefont{G{\"{u}}nther}},
  \bibnamefont{and}
  \bibinfo{author}{\bibfnamefont{J.}~\bibnamefont{Wintterlin}},
  \bibinfo{journal}{Phys. Chem. Chem. Phys.} \textbf{\bibinfo{volume}{10}},
  \bibinfo{pages}{3530} (\bibinfo{year}{2008}).

\bibitem[{\citenamefont{Dil et~al.}(2008)\citenamefont{Dil, Lobo-Checa,
  Laskowski, Blaha, Berner, Osterwalder, and Greber}}]{Dil08}
\bibinfo{author}{\bibfnamefont{H.}~\bibnamefont{Dil}},
  \bibinfo{author}{\bibfnamefont{J.}~\bibnamefont{Lobo-Checa}},
  \bibinfo{author}{\bibfnamefont{R.}~\bibnamefont{Laskowski}},
  \bibinfo{author}{\bibfnamefont{P.}~\bibnamefont{Blaha}},
  \bibinfo{author}{\bibfnamefont{S.}~\bibnamefont{Berner}},
  \bibinfo{author}{\bibfnamefont{J.}~\bibnamefont{Osterwalder}},
  \bibnamefont{and} \bibinfo{author}{\bibfnamefont{T.}~\bibnamefont{Greber}},
  \bibinfo{journal}{Science} \textbf{\bibinfo{volume}{319}},
  \bibinfo{pages}{824} (\bibinfo{year}{2008}).

\bibitem[{\citenamefont{Min et~al.}(2008)\citenamefont{Min, Akbulut,
  Kristiansen, Golan, and Israelachvili}}]{Min08nmat}
\bibinfo{author}{\bibfnamefont{Y.}~\bibnamefont{Min}},
  \bibinfo{author}{\bibfnamefont{M.}~\bibnamefont{Akbulut}},
  \bibinfo{author}{\bibfnamefont{K.}~\bibnamefont{Kristiansen}},
  \bibinfo{author}{\bibfnamefont{Y.}~\bibnamefont{Golan}}, \bibnamefont{and}
  \bibinfo{author}{\bibfnamefont{J.}~\bibnamefont{Israelachvili}},
  \bibinfo{journal}{Nature Mater.} \textbf{\bibinfo{volume}{7}},
  \bibinfo{pages}{527} (\bibinfo{year}{2008}).

\bibitem[{\citenamefont{N'Diaye et~al.}(2006)\citenamefont{N'Diaye, Bleikamp,
  Feibelman, and Michely}}]{Diaye06}
\bibinfo{author}{\bibfnamefont{A.~T.} \bibnamefont{N'Diaye}},
  \bibinfo{author}{\bibfnamefont{S.}~\bibnamefont{Bleikamp}},
  \bibinfo{author}{\bibfnamefont{P.~J.} \bibnamefont{Feibelman}},
  \bibnamefont{and} \bibinfo{author}{\bibfnamefont{T.}~\bibnamefont{Michely}},
  \bibinfo{journal}{Phys. Rev. Lett.} \textbf{\bibinfo{volume}{97}},
  \bibinfo{pages}{215501} (\bibinfo{year}{2006}).

\bibitem[{\citenamefont{Boyen et~al.}(2002)\citenamefont{Boyen, K{\"{a}}stle,
  Weigl, Koslowski, Dietrich, Ziemann, Spatz, Riethm{\"{u}}ller, Hartmann,
  M{\"{o}}ller et~al.}}]{Boyen02}
\bibinfo{author}{\bibfnamefont{H.-G.} \bibnamefont{Boyen}},
  \bibinfo{author}{\bibfnamefont{G.}~\bibnamefont{K{\"{a}}stle}},
  \bibinfo{author}{\bibfnamefont{F.}~\bibnamefont{Weigl}},
  \bibinfo{author}{\bibfnamefont{B.}~\bibnamefont{Koslowski}},
  \bibinfo{author}{\bibfnamefont{C.}~\bibnamefont{Dietrich}},
  \bibinfo{author}{\bibfnamefont{P.}~\bibnamefont{Ziemann}},
  \bibinfo{author}{\bibfnamefont{J.~P.} \bibnamefont{Spatz}},
  \bibinfo{author}{\bibfnamefont{S.}~\bibnamefont{Riethm{\"{u}}ller}},
  \bibinfo{author}{\bibfnamefont{C.}~\bibnamefont{Hartmann}},
  \bibinfo{author}{\bibfnamefont{M.}~\bibnamefont{M{\"{o}}ller}},
  \bibnamefont{et~al.}, \bibinfo{journal}{Science}
  \textbf{\bibinfo{volume}{297}}, \bibinfo{pages}{1533} (\bibinfo{year}{2002}).

\bibitem[{\citenamefont{Wu et~al.}(1994)\citenamefont{Wu, Xu, and
  Goodman}}]{Wu94}
\bibinfo{author}{\bibfnamefont{M.-C.} \bibnamefont{Wu}},
  \bibinfo{author}{\bibfnamefont{Q.}~\bibnamefont{Xu}}, \bibnamefont{and}
  \bibinfo{author}{\bibfnamefont{D.~W.} \bibnamefont{Goodman}},
  \bibinfo{journal}{J. Phys. Chem.} \textbf{\bibinfo{volume}{98}},
  \bibinfo{pages}{5104} (\bibinfo{year}{1994}).

\bibitem[{\citenamefont{Goriachko et~al.}(2007)\citenamefont{Goriachko, He,
  Knapp, Over, Corso, Brugger, Berner, Osterwalder, and Greber}}]{Goriachko07}
\bibinfo{author}{\bibfnamefont{A.}~\bibnamefont{Goriachko}},
  \bibinfo{author}{\bibfnamefont{Y.}~\bibnamefont{He}},
  \bibinfo{author}{\bibfnamefont{M.}~\bibnamefont{Knapp}},
  \bibinfo{author}{\bibfnamefont{H.}~\bibnamefont{Over}},
  \bibinfo{author}{\bibfnamefont{M.}~\bibnamefont{Corso}},
  \bibinfo{author}{\bibfnamefont{T.}~\bibnamefont{Brugger}},
  \bibinfo{author}{\bibfnamefont{S.}~\bibnamefont{Berner}},
  \bibinfo{author}{\bibfnamefont{J.}~\bibnamefont{Osterwalder}},
  \bibnamefont{and} \bibinfo{author}{\bibfnamefont{T.}~\bibnamefont{Greber}},
  \bibinfo{journal}{Langmuir} \textbf{\bibinfo{volume}{23}},
  \bibinfo{pages}{2928} (\bibinfo{year}{2007}).

\bibitem[{\citenamefont{Feidenhans'l}(1989)}]{Feidenhansl89}
\bibinfo{author}{\bibfnamefont{R.}~\bibnamefont{Feidenhans'l}},
  \bibinfo{journal}{Surf. Sci. Rep.} \textbf{\bibinfo{volume}{10}},
  \bibinfo{pages}{105} (\bibinfo{year}{1989}).

\bibitem[{\citenamefont{Bunk et~al.}(2007)\citenamefont{Bunk, Corso, Martoccia,
  Herger, Willmott, Patterson, Osterwalder, van~der Veen, and Greber}}]{Bunk07}
\bibinfo{author}{\bibfnamefont{O.}~\bibnamefont{Bunk}},
  \bibinfo{author}{\bibfnamefont{M.}~\bibnamefont{Corso}},
  \bibinfo{author}{\bibfnamefont{D.}~\bibnamefont{Martoccia}},
  \bibinfo{author}{\bibfnamefont{R.}~\bibnamefont{Herger}},
  \bibinfo{author}{\bibfnamefont{P.~R.} \bibnamefont{Willmott}},
  \bibinfo{author}{\bibfnamefont{B.~D.} \bibnamefont{Patterson}},
  \bibinfo{author}{\bibfnamefont{J.}~\bibnamefont{Osterwalder}},
  \bibinfo{author}{\bibfnamefont{J.~F.} \bibnamefont{van~der Veen}},
  \bibnamefont{and} \bibinfo{author}{\bibfnamefont{T.}~\bibnamefont{Greber}},
  \bibinfo{journal}{Surf. Sci.} \textbf{\bibinfo{volume}{601}},
  \bibinfo{pages}{L7} (\bibinfo{year}{2007}).

\bibitem[{\citenamefont{Oshima and Nagashima}(1997)}]{Oshima97}
\bibinfo{author}{\bibfnamefont{C.}~\bibnamefont{Oshima}} \bibnamefont{and}
  \bibinfo{author}{\bibfnamefont{A.}~\bibnamefont{Nagashima}},
  \bibinfo{journal}{J. Phys. Cond. Matter} \textbf{\bibinfo{volume}{9}},
  \bibinfo{pages}{1} (\bibinfo{year}{1997}).

\bibitem[{Zeg()}]{ZegenhagenThanks}
\bibinfo{note}{The design of the minichamber is based on that of T.-L. Lee and
  J. Zegenhagen of the European Synchrotron Research Facility, whom we
  gratefully acknowledge}.

\bibitem[{\citenamefont{Willmott et~al.}(2005)\citenamefont{Willmott,
  Schlep{\"u}tz, Patterson, Herger, Lange, Meister, Maden, Br{\"o}nnimann,
  Eikenberry, H{\"{u}}lsen et~al.}}]{Willmott05ass}
\bibinfo{author}{\bibfnamefont{P.~R.} \bibnamefont{Willmott}},
  \bibinfo{author}{\bibfnamefont{C.~M.} \bibnamefont{Schlep{\"u}tz}},
  \bibinfo{author}{\bibfnamefont{B.~D.} \bibnamefont{Patterson}},
  \bibinfo{author}{\bibfnamefont{R.}~\bibnamefont{Herger}},
  \bibinfo{author}{\bibfnamefont{M.}~\bibnamefont{Lange}},
  \bibinfo{author}{\bibfnamefont{D.}~\bibnamefont{Meister}},
  \bibinfo{author}{\bibfnamefont{D.}~\bibnamefont{Maden}},
  \bibinfo{author}{\bibfnamefont{C.}~\bibnamefont{Br{\"o}nnimann}},
  \bibinfo{author}{\bibfnamefont{E.~F.} \bibnamefont{Eikenberry}},
  \bibinfo{author}{\bibfnamefont{G.}~\bibnamefont{H{\"{u}}lsen}},
  \bibnamefont{et~al.}, \bibinfo{journal}{Appl. Surf. Sci.}
  \textbf{\bibinfo{volume}{247}}, \bibinfo{pages}{188} (\bibinfo{year}{2005}).

\bibitem[{\citenamefont{Schlep\"utz et~al.}(2005)\citenamefont{Schlep\"utz,
  Herger, Willmott, Patterson, Bunk, Br\"onnimann, Henrich, H\"ulsen, and
  Eikenberry}}]{Schlepuetz05}
\bibinfo{author}{\bibfnamefont{C.~M.} \bibnamefont{Schlep\"utz}},
  \bibinfo{author}{\bibfnamefont{R.}~\bibnamefont{Herger}},
  \bibinfo{author}{\bibfnamefont{P.~R.} \bibnamefont{Willmott}},
  \bibinfo{author}{\bibfnamefont{B.~D.} \bibnamefont{Patterson}},
  \bibinfo{author}{\bibfnamefont{O.}~\bibnamefont{Bunk}},
  \bibinfo{author}{\bibfnamefont{C.}~\bibnamefont{Br\"onnimann}},
  \bibinfo{author}{\bibfnamefont{B.}~\bibnamefont{Henrich}},
  \bibinfo{author}{\bibfnamefont{G.}~\bibnamefont{H\"ulsen}}, \bibnamefont{and}
  \bibinfo{author}{\bibfnamefont{E.~F.} \bibnamefont{Eikenberry}},
  \bibinfo{journal}{Acta Crystallogr. A} \textbf{\bibinfo{volume}{61}},
  \bibinfo{pages}{418} (\bibinfo{year}{2005}).

\bibitem[{\citenamefont{Nagashima et~al.}(1994)\citenamefont{Nagashima, Tejima,
  and Oshima}}]{Nagashima94}
\bibinfo{author}{\bibfnamefont{A.}~\bibnamefont{Nagashima}},
  \bibinfo{author}{\bibfnamefont{N.}~\bibnamefont{Tejima}}, \bibnamefont{and}
  \bibinfo{author}{\bibfnamefont{C.}~\bibnamefont{Oshima}},
  \bibinfo{journal}{Phys. Rev. B} \textbf{\bibinfo{volume}{50}},
  \bibinfo{pages}{17487} (\bibinfo{year}{1994}).

\end{thebibliography}

\end{document}